\documentclass[prb,aps,twocolumn,superscriptaddress,10pt,amsmath,amssymb,nofootinbib,floatfix]{revtex4-1}
\pdfoutput=1
\usepackage{epsfig}
\usepackage{amsmath}
\usepackage{amssymb}
\usepackage{graphicx}
\usepackage{enumitem}
\usepackage{longtable}

\makeatletter 

\begin{document}

\title{Antiferroelectricity in a family of pyroxene-like oxides with
  rich polymorphism}

\author{Hugo Aramberri} \affiliation{Materials Research and Technology
  Department, Luxembourg Institute of Science and Technology, 5 avenue
  des Hauts-Fourneaux, L-4362 Esch/Alzette, Luxembourg}

\author{Jorge \'I\~niguez} \affiliation{Materials Research and
  Technology Department, Luxembourg Institute of Science and
  Technology, 5 avenue des Hauts-Fourneaux, L-4362 Esch/Alzette,
  Luxembourg} \affiliation{Department of Physics and Materials
  Science, University of Luxembourg, 41 Rue du Brill, Belvaux L-4422,
  Luxembourg}

\date{\today}
\begin{abstract}
Antiferroelectrics have potential applications in energy conversion
and storage, but are scarce, particularly among oxides that otherwise
display rich ferroic behaviours. Are we overlooking potential
antiferroelectrics, simply because we have not discovered their
corresponding ferroelectric phase yet? Here we report a
first-principles study suggesting this is the case of a family {\sl
  AB}O$_{3}$ pyroxene-like materials, characterized by chains of
corner-sharing {\sl B}O$_{4}$ tetrahedra, a well-known member being
KVO$_{3}$. The irregular tetrahedra have an electric dipole associated
to them. In the most stable polymorph, the dipoles display an
antipolar pattern with zero net moment. However, upon application of
an electric field, half of the tetrahedra rotate, flipping the
corresponding dipoles and reaching a ferroelectric state.  We discuss
the unique possibilities for tuning and optimization these
antiferroelectrics offer. We argue that the structural features
enabling this antiferroelectric behaviour are also present in other
all-important mineral families.
\end{abstract}
\maketitle

There is currently considerable interest in finding new
antiferroelectric (AFE) materials, owing to their technological
importance and relative
scarcity~\cite{scarceAFEs,rabe2013antiferroelectricity,scarceAFEs2}.
Applications of AFEs rely on their unique response to an applied
electric bias, featuring a double hysteresis loop that is the result
of a field-induced phase transition to a polar state. This double-loop
makes them particularly efficient for energy
applications~\cite{liu2018antiferroelectrics,energyAFE}, as e.g. in
pulsed-power capacitors~\cite{pulsedpowerAFE,chauhan2015anti}. The
field-induced transformation usually results in a large mechanical
response, suitable for transducers and
actuators~\cite{transduceAFE,actuatorAFE}. Finally, AFEs display an
inverse electro-caloric
effect~\cite{Mischenko1270,lu2017large,inverseandconverseEC_AFE}.

To identify new AFEs, one could proceed as follows: Chosen a materials
family and a pertinent high-symmetry structure (e.g., the ideal cubic
phase of perovskite oxides {\sl AB}O$_{3}$), one could use
first-principles simulations to find compositions that
present similarly-strong polar and antipolar phonon instabilities of
this parent phase. Any such compound is likely to present metastable
polar and antipolar polymorphs of similar energy. Then, if the
antipolar state were more stable, and given that an electric field
will always favour the polar one, we would have a good candidate to
display AFE behaviour.

This is exactly the exercise that led to the discoveries here
reported. We ran a high-throughput first-principles study of the
dynamical stability of numerous compounds in the cubic perovskite
phase, and found that alkali vanadates (NaVO$_{3}$, KVO$_{3}$,
RbVO$_{3}$ and CsVO$_{3}$) show dominant and related polar and
antipolar soft modes (see Supp. Note~1 and Suppl. Fig.~S1). Then, when
trying to identify the polar and antipolar polymorphs associated to
these instabilities, we found they result in very large distortions of
the cubic phase, to the point that the perovskite lattice is nearly
destroyed. The resulting energy minima include many low-energy
antipolar polymorphs, as well as their polar counterparts.  Indeed,
according to our calculations, they constitute a very promising family
of AFE compounds.

\begin{figure}[b!]
 \includegraphics[scale=0.05]{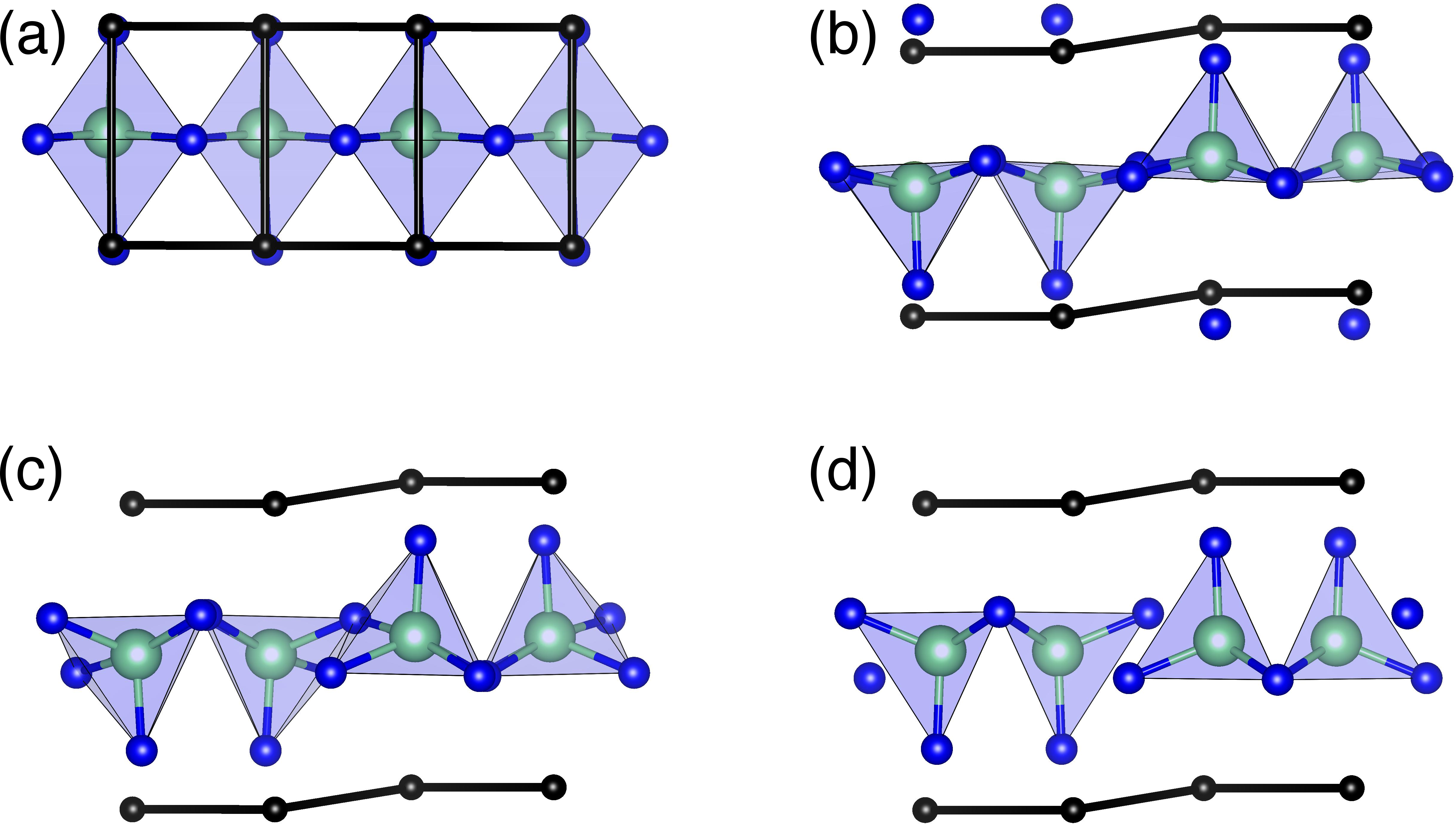}
	\caption{Continuous path connecting the cubic perovskite phase
          ({\bf a}) and the perovskite-derivative antipolar phase
          ({\bf d}). The {\sl A}, {\sl B} and O atoms are shown in
          black, green and blue, respectively. Initially one B-O link
          per perovskite unit cell is broken due to a transition to a
          supertetragonal state ({\bf b}), while a rotation of the
          resulting pyramids results in the rupture of a second B-O
          bond per cell (transition from {\bf c} to {\bf d}).}
	\label{fig1}
\end{figure}

{\sl Results.--} In order to identify stable structures, we run
first-principles molecular dynamics simulations (see Methods) of
several alkali vanadates, which reveal the existence of numerous
low-energy metastable polymorphs with some common features: a strong
tetragonal distortion, with $c/a$ ratios larger than 1.3; a sublattice
of {\sl A} cations that retain a perovskite-like configuration; and a
4-fold coordination of the vanadium atoms yielding corner-sharing
VO$_{4}$ tetrahedra.

The connection between the perovskite phase and the ground state of
these vanadates is sketched in Fig.~\ref{fig1}, which shows the result
of a nudged elastic band (NEB) calculation (see Methods for details).
Starting from the perovskite structure (Fig.~\ref{fig1}a), the system
undergoes a transition to a phase with a strong tetragonal distortion,
in which a first V-O bond is broken in each {\sl A}VO$_{3}$ unit
(Fig.~\ref{fig1}b, and Suppl. Video~\#1). The obtained structures are
reminiscent of supertetragonal phases known for perovskites like
BiFeO$_{3}$ and PbVO$_{3}$, which show a large $c/a$ ratio and
consequently a layered structure formed by BO$_{5}$
pyramids~\cite{BFOsupertetraDFT,synthesisPbVO3}.  In a subsequent
step, the VO$_{5}$ square pyramids rotate along the V-O bond of the
apical oxygen (Fig.~\ref{fig1}c), while the V cations move towards one
of the oxygen atoms at the base of the pyramid (Fig.~\ref{fig1}d),
ending in the rupture of a second V-O bond. Each of the resulting
VO$_{4}$ groups is corner-linked to two adjacent tetrahedra, forming
chains.

\begin{figure}[t!]
 \includegraphics[width=0.95\linewidth]{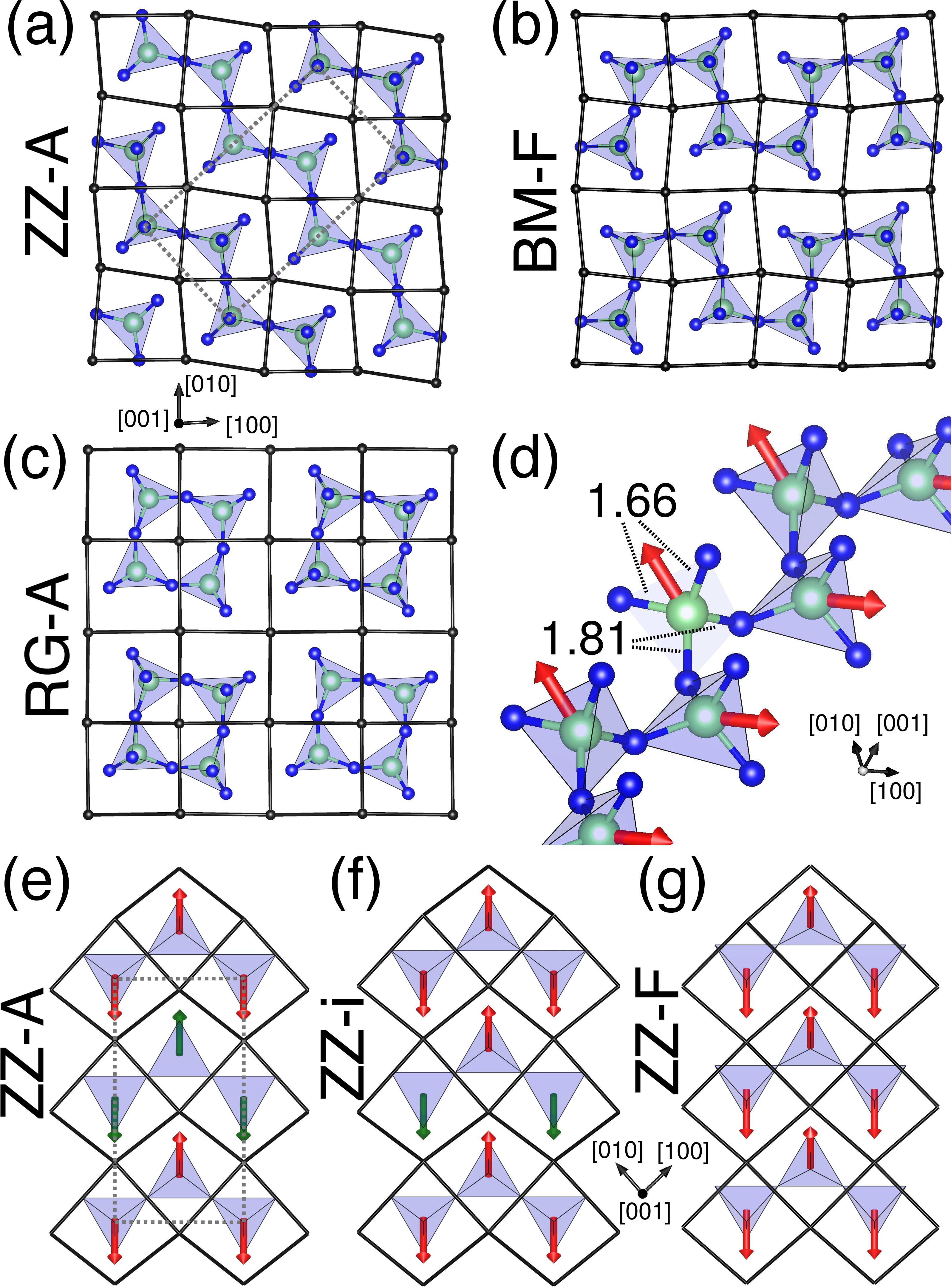}
	\caption{Representative cases of the pyroxene-like stable polymorphs
	found. (a) and (b) depict two polymorphs with 1D O tetrahedra chains,
	with a ZZ (ground state) and a BM structure, respectively. An example
	of a polymorph with closed (0-dimensional) O tetrahedra structures is
	shown in (c) (RG-A). The inter-chain coupling of the dipoles is
	indicated by the letters A (antiferroelectric) and F (ferroelectric) in
	the structure names. The microscopical origin of the local dipoles
	(red arrows) is shown in (d), where the indicated V-O bond distances
	are given in \AA\ for KVO$_{3}$. The ZZ-A, ZZ-i and ZZ-F states are
	shown schematically in panels (e), (f), and (g), respectively.
	}\label{fig2}
\end{figure}

We obtain several polymorphs featuring the mentioned vertex-sharing
oxygen tetrahedra, inspired by metastable phases observed in our
molecular dynamics simulations. The tetrahedra can form
one-dimensional (1D) zigzag (ZZ) structures along the [110]
pseudo-cubic direction (Fig.~\ref{fig2}a). The chains can
alternatively follow a 1D battlement-like (BM) pattern along the [010]
axis (Fig.~\ref{fig2}b), or even form closed loops yielding
zero-dimensional ring (RG) structures (Fig.~\ref{fig2}c). In the ZZ
structure the apical oxygens point in opposite directions for adjacent
chains, closely resembling what is found in inosilicate
pyroxenes~\cite{burnham1967proposed,hawthorne1977crystal,putnis1992introduction},
while the RG structures have close analogs in the cyclosilicate
family~\cite{zoltai1960,buerger1960relative,COLIN1993242}. The ZZ
structure is the most stable one, the lowest-energy BM and RG
structures lying 117~meV and 125~meV \textit{per} formula unit (f.u.)
above. The ZZ ground state we find for KVO$_{3}$ (as well as for
RbVO$_{3}$ and CsVO$_{3}$) is orthorhombic with space group $Pbcm$ and
coincides with the experimentally observed
structure~\cite{KVOorig1954,Petrasova1958,KVOslovak1958,evans1960crystal,hawthorne1977crystal,RamanKVORVOCVO,de1992vib}.

Importantly, the V-O bonds forming the backbone of the tetrahedra
chains are longer than the remaining two hovering V-O
bonds~\cite{hawthorne1977crystal,angsten2018electronic} -- see bond
lengths for the ZZ polymorph of KVO$_{3}$ in Fig.~\ref{fig2}d. This
difference of bond lengths induces a local electric dipole which lies
along the direction defined by the V cation and the centre of the
tetrahedron edge formed by the lingering oxygens, as shown in
Fig.~\ref{fig2}d.

We now focus on the ZZ phase of KVO$_{3}$ since it is the ground state
of the best-established compound among those studied. For a given
chain, the ZZ structure gives rise to an antipolar pattern of the
in-plane component of the dipoles, while the out-of-plane component
remains constant. Since the apical oxygens point in opposite
directions for contiguous chains, the out-of-plane component changes
sign from chain to chain, yielding a striped antipolar pattern
(Fig.~\ref{fig2}e). An obvious question arises: is KVO$_{3}$
antiferroelectric? A positive answer can only be given if a related
ferroelectric (FE) phase accessible via an electric field is found.

\begin{figure}[t!]
 \includegraphics[scale=0.85]{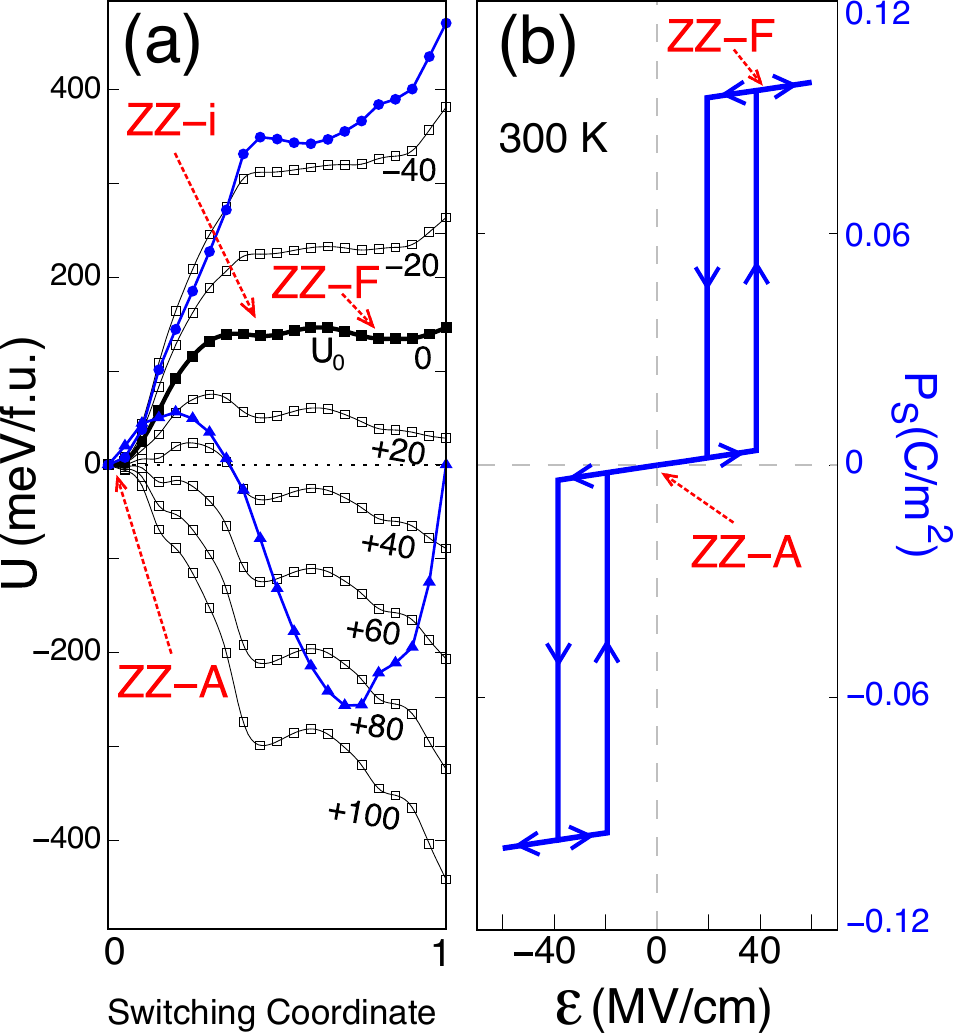}
	\caption{(a) Minimum energy path for switching between the ZZ-A
	and the ZZ-F states in KVO$_{3}$ as obtained with the NEB method
	(solid thickest black line). Blue lines correspond to the spontaneous
	polarisation components out-of-plane (circles) and in-plane,
	perpendicular to the chains (triangles). The vertical axis for the
	polarisation is shared with panel (b) and at the right of the latter.
	The black curves show the energy density profiles under different
	perpendicular electric fields, being the value of the fields indicated
	on each line in MV/cm. (b) Estimated double hysteresis loop at 
	$T = 300$~K.
	}\label{fig3}
\end{figure}
 
We crucially realised that a rotation of a tetrahedron about the
backbone edge is tantamount to switching the out-of-plane component of
its electric dipole. We thus constructed a ZZ structure with FE
inter-chain ordering (ZZ-F) that turned out to be stable and
135~meV/f.u. higher in energy than the antipolar ground state
(ZZ-A). The polar order of the ZZ-F polymorph is shown schematically
in Fig.~\ref{fig3}g. We calculate its spontaneous
polarisation~\cite{modernTpol} to be 0.093~C/m$^2$ (0.054~C/m$^2$) for
the out-of-plane (in-plane) component, of the same order of magnitude
as perovskite FEs like BaTiO$_{3}$ (0.43~C/m$^2$)~\cite{BTOVanderbilt}.
Since the V-O distances in the ZZ-A
phase differ by less than 0.5~\% from those of the ZZ-F phase, we
estimate the out-of-plane sublattice polarisation in the ZZ-A phase to
be approximately 0.046~C/m$^2$, i.e., half of the ZZ-F polarisation
value. Using the tools described in Methods, we find that the
optimized ZZ-F state has $Pm$ symmetry. An intermediate ferrielectric
state (ZZ-i), with half of the tetrahedra in one of the chains flipped
(Fig.~\ref{fig3}f), was also found to be metastable for KVO$_{3}$.

What remains in order to confirm the AFE character of KVO$_{3}$ is to
find a connecting path for the field-induced transition. To this end
we carry out NEB calculations between the ZZ-A and the ZZ-F
polymorphs. The results are shown in Fig.~\ref{fig3}a (thick black
line). A continuous energy path is found with an energy barrier of
147~meV/f.u. The switching of the dipole chain occurs in a step-wise
fashion, with half of the tetrahedra rotating at a first stage -- and
ending up in the ZZ-i phase --, and the remaining ones rotating at a
second stage (see Suppl. Video \#2). Note that we extend the NEB path
up to a FE state in which the polarisation lies fully out-of-plane;
for KVO$_{3}$ such a state, with space group $Pma2$, is a saddle point
of the energy. The two components of the spontaneous polarisation that
acquire non-zero values along the switching path are shown in blue
solid lines in Fig.~\ref{fig3}a. The out-of-plane polarisation
increases abruptly in the first stage, and slower in the second one.
Along the path a non-zero in-plane polarisation develops,
perpendicular to the tetrahedron chains, showing an oscillating
behaviour due to the rigid dipoles crossing the plane of the chains in
each of the two switching steps.

Finally, we estimate the behaviour of KVO$_{3}$ under application of
an out-of-plane electric field. The response to the applied field can
be approximated by constructing an electric enthalpy with the form
\begin{equation}
  U=U_{0}-v\boldsymbol{\cal E}\cdot \boldsymbol{P} \; ,
\end{equation}
where $U_{0}$, $\boldsymbol{P}$ and $v$ are, respectively, the energy,
polarisation and volume at zero field at each step of the path, as
obtained from first principles; $\boldsymbol{\cal E}$ is the applied
electric field. The ZZ-A state is chosen as the zero of energy for
convenience. By introducing this approximated enthalpy we avoid
running costly first-principles calculations explicitly considering an
applied field. In Fig.~\ref{fig3}a the energy profile of KVO$_{3}$
along the switching path is shown for different values of the field.

The hysteresis loop of the polarisation under an out-of-plane electric field
can be numerically reconstructed from the enthalpies and polarisations in
Fig.~\ref{fig3}a. More specifically, we consider the case of $T = 300$~K,
assuming a thermal energy of 26~meV/f.u., and obtain the results in
Fig~\ref{fig3}b. (See Supp. Note~2 and Supp. Fig.~S2 for details on the
calculation of these loops.) The AFE$\rightarrow$FE switching, and the
FE$\rightarrow$AFE back-switching, occur when the applied field lowers the
corresponding energy barrier below the thermal activation energy. At 300~K,
this occurs for 40~MV/cm (switching) and 20~MV/cm (back-switching), and we
observe a sizeable hysteresis. The efficiency of the material as an AFE
capacitor would be about 50~\%. 

The computed switching fields are very large compared to those that
can be applied experimentally in similar oxides before inducing
leakage (i.e., about 1~MV/cm). To understand the implications of this
result, let us first note that first-principles estimates like ours
are known to exaggerate FE coercive fields by factors of up to two
orders of magnitude~\cite{Lisenkovoverestimate,Daumontoverestimate},
probably because they miss effects (e.g., easier nucleation of the
field-induced phase at defects) that play an important role at
controlling the transformation kinetics. However, in the case of a
AFE$\leftrightarrow$FE transformation, the coercive bias must be as
large as to equalize the energies of the polar and antipolar states.
Our simulations do suggest that a very strong field (of about
28~MV/cm) is needed to achieve this in KVO$_{3}$; hence,
notwithstanding possible inaccuracies in our estimate, it seems
unlikely KVO$_{3}$ can be experimentally switched. Having said this,
as we discuss below, we have reasons to believe that there are
promising ways to optimize the switching characteristics of KVO$_{3}$
and related compounds, e.g. by means of appropriate chemical
substitutions, considerably reducing the fields required to achieve
AFE behaviour.

{\sl Discussion.--} Let us start by commenting on three possibilities
for AFE optimization and tuning that these materials offer.  First,
our simulations indicate that the behaviour of RbVO$_{3}$ and
CsVO$_{3}$ is very similar to that of KVO$_{3}$. The ZZ-A phase is
also the ground state for these compounds, and the ZZ-F phase is a
metastable polymorph with an out-of-plane spontaneous polarisation of
0.078~C/m$^2$ and 0.052~C/m$^2$, respectively. The decrease in the
polarisation, as compared to the result for KVO$_{3}$, can be ascribed
to the larger unit cell volume. The ZZ-F state of RbVO$_{3}$ shows a
non-negligible in-plane component (0.037~C/m$^2$), while for
CsVO$_{3}$ the polarisation is fully out-of-plane.  The calculated
energy barrier between the ZZ-A and ZZ-F states decreases with
increasing size of the {\sl A} cation (being of 128~meV/f.u. and
112~meV/f.u. for RbVO$_{3}$ and CsVO$_{3}$, respectively). Further,
the energy difference between the ZZ-F and the ZZ-A states follows the
same trend, being of 117~meV/f.u.and 97~meV/f.u. for RbVO$_{3}$ and
CsVO$_{3}$, respectively (see Suppl. Note~3 and Suppl. Fig.~S3).

Second, regarding the possibility of having different {\sl B} cations,
further calculations indicate that ZZ-A is a metastable structure for
all the alkali tantalates and niobates, in particular for the
well-studied FE KNbO$_{3}$. Therefore, a morphotropic phase boundary
(MPB) between the FE perovskite phase and the ZZ-A phase must exist
for the solid mixture KV$_{1-x}$Nb$_{x}$O$_{3}$. Our preliminary
studies indicate that such an MPB occurs between $x$=0.375 and
$x$=0.5. At such compositions, the energies of the AFE (ZZ-A) and FE
(KNbO$_{3}$-like) states become very close (cross). Hence, these solid
solutions naturally provide us with AFE and FE states that are very
close in energy -- thus solving the main difficulty mentioned above to
obtain AFE$\leftrightarrow$FE switching at moderate fields --, and
seem ideal candidates to yield AFE materials with optimized
properties.

Third, as already mentioned, we find polymorphs with different
tetrahedral arrangements, forming BM and RG patterns. Interestingly,
for the three alkali vanadates considered, the most stable dipolar
order is FE in the BM case. Moreover, the (shear) strains are distinct
for the ZZ, BM and RG geometries (see the different structures of the
{\sl A} cation sublattice in Fig.~\ref{fig2}).  Hence, different chain
arrangements, and in turn the polar order, may be accessible by
growing these materials on appropriate substrates that impose suitable
epitaxial conditions. Further, as can be appreciated from
Figs.~\ref{fig2}e and \ref{fig2}g, the ZZ-F state is more {\sl square}
in-plane than the ZZ-A, which suggests that a suitable substrate may
allow us to tune the corresponding energy difference.

It is also worth noting some chemical and structural aspects of the
compounds studied in this work. The origin of the peculiar polymorphs
found seems to be the small size of the V$^{5+}$ cation (nominally,
0.54~\AA\ for V$^{5+}$ in a 6-fold coordination, and 0.355~\AA\ in a
4-fold coordination)~\cite{Shannon} relative to the O$^{2-}$ ionic
radius (1.35~\AA)~\cite{Shannon}.  The size difference is such that
all vanadates lie below the octahedral limit~\cite{Filip5397}, which
is known to lead to lower {\sl B}-cation coordination in
perovskites~\cite{goldschmidt1926gesetze,megaw1973crystal,Filip5397}.
Further, we find that the V-O bond lengths change by less than 1~\%
across the switching process between the ZZ-A and ZZ-F phases.  In
fact, the deviations in V-O bond lengths among all the ZZ, BM, and RG
polymorphs of KVO$_{3}$ are below 1~\%, and even the differences among
the three studied vanadates remain below 1~\%. The V-O bonds, and
consequently the local dipoles, thus prove to be very
rigid~\cite{selbin1965chemistry}, suggesting that in these materials
phase transitions involving dipoles will probably be of the
order-disorder type~\cite{perez2000displacive}.  More importantly,
this bond stiffness also ensures that the dipoles will not vanish;
therefore, these compounds can be viewed as model AFEs whose behaviour
is analogous to that of antiferromagnets~\cite{Kittel}.

Also of note is the switching in these compounds -- by quasi-rigid
rotations of VO$_{4}$ molecular-like groups --, which is rather unique
as compared to similar transformations in inorganic FE and AFE
materials. Indeed, the identified switching path suggests that, in
these materials, such transformations will typically occur through
many steps. The present calculation shows a switching in \textit{only}
two steps, which seems a direct consequence of the finite size of the
simulation box employed; however, larger simulation cells would
probably reveal a many-step process. This is strongly reminiscent of
memristors, in which the electric resistance can be tuned
quasi-continuously by taking advantage of a controllable multi-step
transformation. Therefore, our results suggest that pyroxene-like AFEs
could find application in memristor
devices~\cite{chanthbouala2012ferroelectric}.

Finally, the findings here reported hint at a promising strategy to
discover further AFE materials.  Pyroxenes,
pyroxenoids~\cite{pyroxenoidsbuerger1956,pyroxenoidmamedov1956} and
many other {\sl AB}O$_{3}$ compounds -- like e.g. the recently
synthesised BiGaO$_{3}$~\cite{belik2006high} or the vanadates
NH$_{4}$VO$_{3}$~\cite{Lukesh1950,evans1960crystal},
TlVO$_{3}$~\cite{ganne1974structure},
NaVO$_{3}$~\cite{SorumNaVO,MarumoNaVO,miller1962nb1}, 
$\alpha$-AgVO$_{3}$~\cite{kittaka1999crystal}, and 
LiVO$_{3}$~\cite{shannon1973crystal}, with structures akin to that of
KVO$_{3}$ -- also contain chains of irregular oxygen tetrahedra. Since
such tetrahedra host an electric dipole, these compounds can be viewed
to present an antipolar ground state, and are candidates to display
AFE behaviour. The situation is similar to that of BiVO$_{4}$, an
extensively investigated material (for its catalytic
properties~\cite{yu2006effects}) that is formed by irregular VO$_{4}$
tetrahedra whose corresponding electric dipoles order in an antipolar
pattern, and which has recently been proposed as a possible
AFE~\cite{toledano2016theory}. Hence, we hope the present work will
stimulate experimental and theoretical activities to explore this
intriguing possibility, namely, that some of the best-known and
most-abundant minerals on Earth may be antiferroelectrics in disguise!

{\bf Acknowledgements}. Work funded by the Luxembourg National
Research Fund through the project
INTER/ANR/16/11562984/EXPAND/Kreisel. We would like to thank Enric
Canadell (ICMAB-CSIC) for his valuable comments and for pointing to
Ref.~\onlinecite{angsten2018electronic}

{\bf Author Contributions}. Work carried out by H.A. and supervised by
J.\'I.

{\bf Methods}. In this work we employ first-principles calculations
within the density functional theory (DFT) framework as implemented in
the Vienna {\it Ab-initio} Simulation Package
(\textsc{vasp})~\cite{VASP1,VASP2} to obtain the crystal structure and
relative energies of the different polymorphs and compounds. The
plane-wave cut-off for the basis set is set to 500~eV in all cases. We
choose the Perdew-Burke-Ernzerhof functional modified for solids
(PBEsol)~\cite{PBEsol} as the exchange-correlation. For perovskite
primitive cell calculations (5-atom cell) a $6\times 6\times 6$
Monkhorst-Pack~\cite{monkhorstpack} grid is employed for sampling the
Brillouin zone, which yields well-converged results for the three
alkali vanadates. The primitive cell of the ZZ-A ground state of
KVO$_{3}$, RbVO$_{3}$ and CsVO$_{3}$ is a $\sqrt{2}\times
2\sqrt{2}\times 1$ supercell with respect to the ideal perovskite; for
such simulation cells we employ a $4\times 3\times 6$ k-point
sampling. In all structural relaxations the crystals are allowed to
fully relax until atomic forces become smaller than 0.01~eV/\AA. The
molecular dynamics calculations are carried out within the canonical
ensemble using a Langevin thermostat, as implemented in \textsc{vasp}. A
$2\times 2\times 2$ supercell is employed for this purpose (with
respect to the 5-atom cell perovskite), and the k-sampling is reduced
to $2\times 2\times 2$ to speed up the calculations. A wide range of
temperatures is studied, from 10~K to 1000~K.

The space groups of the studied crystal structures are determined
employing the \textsc{spglib} library through its implementation in
the \textsc{phonopy} package~\cite{phonopy}.

We used the \textsc{vesta} visualization package~\cite{VESTA}
to prepare some of the Figures and Supplementary Videos.

The calculations of the switching energy barrier and the connection
between the perovskite structure and the ZZ-A state are obtained
through the NEB method~\cite{NEBorig} along with the climbing-NEB
(cNEB)~\cite{henkelman2000climbing} modification as implemented by the
Henkelman group in the Virtual Transition State Theory (\textsc{vtst}) tools
package for \textsc{vasp}~\cite{henkelman2000climbing}. A total of 19 (7)
images are employed in the NEB calculations of the switching energy
barrier (connection to the perovskite structure).

%

\end{document}